\documentclass[twocolumn,showpacs,amsmath,amssymb,pre]{revtex4}

\usepackage{graphicx}

\begin{document}

\title{Influence of friction on granular segregation}

\author{Stephan Ulrich}
\altaffiliation[Present address: ]{Institut f\"ur Theoretische
Physik, Georg-August-Universit\"at G\"ottingen, 37073 G\"ottingen,
Germany}
\author{Matthias Schr\"oter}
\email{schroeter@chaos.utexas.edu}
\author{Harry L. Swinney}
\affiliation{Center for Nonlinear Dynamics and Department of
Physics, The University of Texas at Austin, Austin,
         Texas 78712, USA}
\date{\today}

\begin{abstract}
Vertical shaking of a mixture of small and large beads can lead to
segregation where the large beads either accumulate at the top of
the sample, the so called Brazil Nut effect (BNE), or at the
bottom, the Reverse Brazil Nut effect (RBNE). Here we demonstrate
experimentally a sharp transition from the RBNE to the BNE when
the particle coefficient of friction increases due to aging of the
particles. This result can be explained by the two competing
mechanisms of buoyancy and sidewall-driven convection, where the
latter is assumed to grow in strength with increasing friction.
\end{abstract}


\maketitle Segregation of particles in granular media is a common
problem in the chemical and pharmaceutical industries and in
materials processing. Several physical mechanisms have been found
to lead to segregation \cite{kudrolli:04,schroeter:06}, but their
dependence on material parameters and especially the role of
friction \cite{srebro:03,pohlman:06,pica_ciamarra:06} are not well
understood. Here we report experimental results on the segregation
of a vertically shaken mixture of 2.4 mm diameter brass spheres
and 1.4 mm diameter glass spheres. Samples of new particles
exhibit a RBNE where all of the large particles accumulate at the
bottom plate of the sample. However, after about 25 hours of
continuous shaking a sharp transition to a BNE occurs for
unchanged driving conditions. Measurements show that the main
effect of the aging is an increase of the coefficient of friction
$\mu$. The initial state (RBNE) can be restored with an aggressive
cleaning procedure which also decreases $\mu$. The cleaned sample
shows the same transition to BNE during subsequent aging.

\begin{figure}[t]
  \begin{center}
    \includegraphics[angle=270,width=8.5cm]{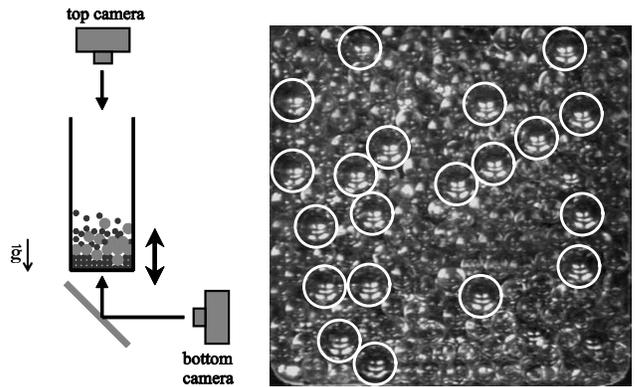}
    \caption{Left: Experimental Setup. The particles are shaken vertically in a square glass tube.
        Pictures are  taken from the top and  the bottom (using a mirror) of the
        granular sample.
      Right: An image of the bottom plate. All large particles detected by the image processing
        are marked with white circles. }
    \label{setup}
  \end{center}
\end{figure}

{\it Experiment.-} The experimental setup is shown in
Fig.~\ref{setup}. The particles are shaken in a 21\;mm x 21\;mm
square glass tube with a height of 150\;mm.  The cell is mounted
on an electromechanical shaker (Vibration Test Systems VG100c) and
shaken sinusoidally with an amplitude $A$ and a fixed frequency $f
= 20$ Hz. The dimensionless shaking acceleration is $\Gamma = (2
\pi f)^2 A / g$, where $g$ is the acceleration due to gravity.
The sample consists of 74 brass (alloy 260) beads with 2.4 mm
diameter ($\rho$ = 8.4 g/cm$^3$) and about 820 glass beads
(Ceroglass GSR 14) with 1.4 mm diameter ($\rho$ = 2.5 g/cm$^3$).

Measurements of each segregation state were performed at both
$\Gamma$ = 2 and 5. Before each measurement the beads were shaken
for 2 minutes at the $\Gamma$ of the measurement; this time was
found to be sufficient in \cite{schroeter:06} for the system to
reach a stationary state, independent of the previous shaking
history. The segregation state for each condition was determined
by stopping the shaker and obtaining images from the top and
bottom surface of the sample. All large particles visible in these
images were then detected with image processing, allowing a
quantitative characterization of the segregation (for details see
\cite{schroeter:06}). After each set of measurements the beads
were aged for one hour by shaking them constantly at $\Gamma$ = 5
before making a new set of measurements.

Table \ref{tab:friction} shows that the aging of the particles is
accompanied by an  increase of $\mu$. Treating the particles for
10 minutes in an ultrasonic bath with a concentrated solution of
Alconox powdered cleaner reduces $\mu$.
\begin{table}
\begin{tabular} {l|c|c}
 &  ``cleaned'' particles  &  aged particles\\
\hline
on a brass chute:&&\\
\hspace{3mm} brass sphere &  ${0.169 \pm 0.016}$ &  ${0.196 \pm 0.009}$\\
\hspace{3mm} glass sphere &  ${0.151 \pm 0.003}$ &  ${0.162 \pm 0.006}$ \\
\hline
on a glass chute: &&\\
\hspace{3mm} brass sphere &  $0.086 \pm 0.009$ & $0.111 \pm 0.003$ \\
\hspace{3mm} glass sphere &  $0.089 \pm 0.006$ & $0.095 \pm 0.004$ \\

\end{tabular}
\caption{Static coefficients of friction  $\mu$ of glass and brass
spheres that were either freshly surface treated or  aged by
shaking for 28 hours. Measurements were performed with three
particles glued to the bottom of a small toboggan. $\mu$ was
determined from the angle when the toboggan started to slide on a
brass or glass chute. This method has been found to yield similar
results to measurements using a tribometer or based on the angle
of repose \cite{pohlman:06}. Each value reported here is the
average of five different sets of balls.} \label{tab:friction}
\end{table}

To exclude alternative causes for the observed transition we also
measured the coefficient of restitution $e$ by dropping the
particles from a height of 30 cm on a plate of the same material.
For both cleaned and aged particles we measured $e= 0.65 \pm 0.02$
for the brass and   $e = 0.89 \pm 0.01$ for the glass particles.
The relative humidity (measured with a Honeywell HIH 3610 sensor)
was 54 $\pm$ 3\% during the experiment and the fluctuations were
not correlated with the transition. Cleaning the container and
particles with soap and water did not change the BNE back to a
RBNE. Also, pausing the aging process for up to 48 hours did not
influence the transition.

{\it Results and discussion.-}
\begin{figure*}[h]
  \begin{center}
    \includegraphics[width=16cm]{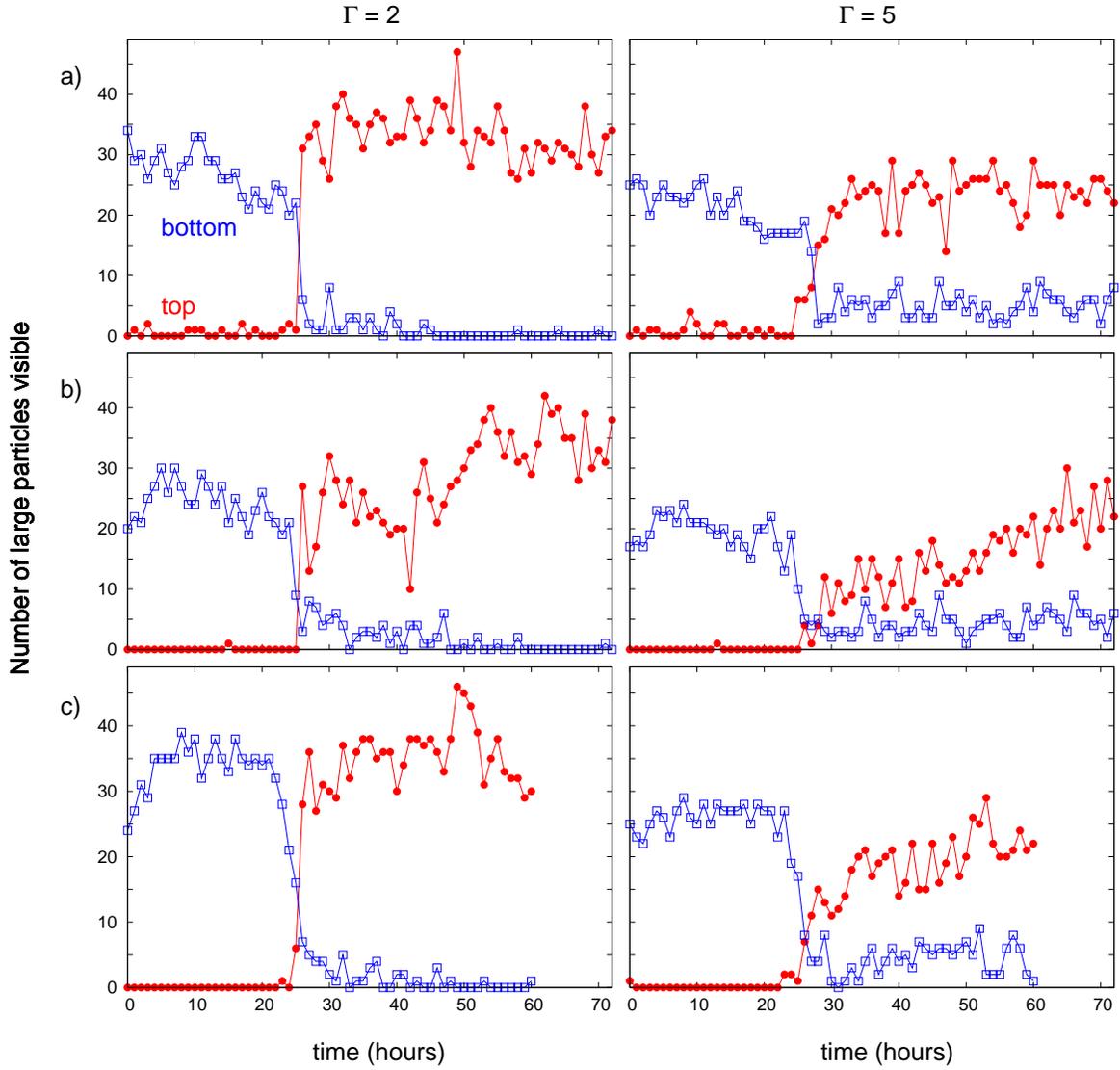}
    \caption{(Color online). Numbers of large particles at the top and bottom surface of the sample
    measured at $\Gamma$ = 2 (left column) or 5 (right column).
        In both cases the segregation changes from a Reverse Brazil Nut state to a Brazil Nut state
    after about 25 hours of aging the particles. Experiment (a) started with new particles;
    the particles in (b) and (c) were treated in
    an alkaline ultrasound bath before the start of the experiment.
    These data were obtained for $f = 20$ Hz.
    }
    \label{fig:exp}
  \end{center}
\end{figure*}
Figure \ref{fig:exp} (a) shows for new particles a sharp
transition from the RBNE state to the BNE state after 25 hours
both at $\Gamma$ = 2 and 5. Continued shaking up to 180 hours did
not change the BNE observed. However, Fig. \ref{fig:exp}(b) and
(c) demonstrate that the surface treatment described above resets
the initial RBNE. Also the transition to the BNE state after about
25 hours is reproducible.

This transition can be explained by the interaction of two
segregation mechanisms: buoyancy/weight
\cite{shinbrot:98,shishodia:01,trujillo:03,huerta:04,gutierrez:04,pica_ciamarra:06}
and sidewall-driven convection
\cite{knight:93,duran:94,poeschel:95,lan:97,shinbrot:98,moebius:05}.
Initially the dense brass particles sink to the bottom of the
vibration-fluidized sample due to their buoyancy being smaller
than their weight \cite{huerta:05}. This mechanism does not depend
strongly on the frictional properties of the beads as evidenced by
its existence in simulations with frictionless particles
\cite{shishodia:01}.

The second segregation mechanism present in our setup and leading
on its own to a BNE is sidewall-driven convection: the difference
in packing density during the upward and downward motion of the
container gives rise to a convection roll going downwards at the
sidewalls and upwards in the center. Segregation occurs because
the larger particles move upwards in the center but have a lower
probability of entering the much thinner downstream layer at the
container walls. An increase in $\mu$ can be expected to
strengthen the convective motion \cite{knight:97,grossman:97} and
therefore also strengthen this segregation mechanism.

The transition at $\Gamma$ = 2 is steeper than the transition at
$\Gamma$ = 5. This can be explained by the increasing importance
of a third mechanism, called thermal diffusion. Thermal diffusion
describes the tendency of the larger particles to accumulate in a
minimum of the granular temperature; its strength increases with
increasing $\Gamma$ and leads to a RBNE state in shallow layers
like the ones considered here \cite{schroeter:06,serero:06,garzo:06}.

{\it Conclusion.-} The transition from RBNE to BNE can be
explained by assuming that during the aging process $\mu$ and
consequentially the sidewall-driven convection increase until this
mechanism surmounts the initially dominant buoyancy/weight
mechanism.  The sharpness of the observed transition indicates
that minute changes in the coefficient of friction can have
dramatic effects on the segregation behavior. This insight could
be important for the industrial handling of granular matter.

{\it Acknowledgments.-} We thank W.D. McCormick and Yair Shokef
for helpful discussions. This work was supported in part by Robert
A. Welch Foundation grant F-0805.


\end{document}